\documentclass[lettersize,journal]{IEEEtran}
\usepackage{amsmath,amsfonts}
\usepackage{algorithmic}
\usepackage{algorithm}
\usepackage{array}
\usepackage[caption=false,font=normalsize,labelfont=sf,textfont=sf]{subfig}
\usepackage{textcomp}
\usepackage{stfloats}
\usepackage{url}
\usepackage{url}
\usepackage{verbatim}
\usepackage{graphicx}
\usepackage{cite}
\usepackage{amsthm}
\newtheorem{theorem}{Theorem}[section]
\usepackage[dvipsnames]{xcolor}
\usepackage{enumerate}

\usepackage[T1]{fontenc}
\usepackage[utf8]{inputenc}
\usepackage{tabularx,ragged2e,booktabs,caption}
\newcolumntype{C}[1]{>{\Centering}m{#1}}

\begin{document}

\title{Spectral Gap-Based Seismic Survey Design}

\author{Oscar L\'opez$^{1}$, Rajiv Kumar$^{2}$, Nick Moldoveanu$^3$ and Felix J. Herrmann$^4$ 
\\$^1$Harbor Branch Oceanographic Institute, Florida Atlantic University
\\$^2$Schlumberger WesternGeco
\\$^3$Independent Geophysical Consultant
\\$^4$School of Earth and Atmospheric Sciences, Georgia Institute of Technology}


\maketitle

\begin{abstract}
Seismic imaging in challenging sedimentary basins and reservoirs requires acquiring, processing, and imaging very large volumes of data (tens of terabytes). To reduce the cost of acquisition and the time from acquiring the data to producing a subsurface image, novel acquisition systems based on compressive sensing, low-rank matrix recovery, and randomized sampling have been developed and implemented. These approaches allow practitioners to achieve dense wavefield reconstruction from a substantially reduced number of field samples.

However, designing acquisition surveys suited for this new sampling paradigm remains a critical and challenging role in oil, gas, and geothermal exploration. Typical random designs studied in the low-rank matrix recovery and compressive sensing literature are difficult to achieve by standard industry hardware. For practical purposes, a compromise between stochastic and realizable samples is needed. In this paper, we propose a deterministic and computationally cheap tool to alleviate randomized acquisition design, prior to survey deployment and large-scale optimization. We consider universal and deterministic matrix completion results in the context of seismology, where a bipartite graph representation of the source-receiver layout allows for the respective \textit{spectral gap} to act as a quality metric for wavefield reconstruction. We provide realistic scenarios to demonstrate the utility of the spectral gap as a flexible tool that can be incorporated into existing survey design workflows for successful seismic data acquisition via low-rank and sparse signal recovery.
\end{abstract}


\section{Introduction}
\label{introduction}
Seismic exploration is a complex and lengthy process involving many expensive components. To ensure high quality imaging in frugal manner, survey designs are an important initial step that determine the efficiency of data acquisition before costly in-field deployment \cite{survey1,survey2,survey3}. The typical components for data acquisition survey designs are:
\begin{enumerate}
    \item Definition of seismic survey objectives in terms of reservoir depth, type of reservoir, required vertical and horizontal imaging resolution.
    \item Analysis of the existent seismic, well, geological data, near surface and environmental conditions in the survey area.
    \item Seismic modeling and imaging of different acquisition scenarios to evaluate placement of sources and receivers based on resolution, signal-to-noise requirements, and inherent noise levels.
    \item Selection of optimum acquisition parameters based on modeling, imaging results, and the cost of acquiring and processing the data.
\end{enumerate}
In all steps, it is essential for a survey design study to have a realistic earth model of the area. If geological and geophysical information in the survey area do not allow an earth model to be built, existing models available in the industry can be informative when the imaging challenges of the survey area are comparable. For instance, the SEG Seam model is quite often use for survey design studies to evaluate the acquisition geometries for sub-salt exploration \cite{NickSEG}.

To further complicate the survey design process, recent years have introduced a new paradigm for data acquisition due to the advent of compressive sensing (CS) \cite{CS}. Low-rank matrix recovery (LRMR) and CS based seismic exploration has successfully materialized from an academic novelty to an indispensable industrial tool \cite{SMC,SMC2,SMC3,SMC4}. As a popular example, seismic data reconstruction from limited measurements provides an instance of the \emph{matrix completion} problem \cite{MC,MC2,MC3,MC4}. These approaches allow for simultaneous acquisition and compression of seismic data, where a substantially reduced number of field samples allow dense wavefield reconstruction via large-scale optimization. As a consequence, practitioners may achieve high-fidelity imaging and post-processing with economical surveys. 

While presenting many advantages, LRMR and CS-based acquisition remains a challenging task with many open questions in addition to the pre-existing challenges of Nyquist-based survey design. In particular, it is not clear how to fulfill step 3) above in an practical manner with respect to LRMR and CS. In the theory of matrix completion, generating appropriate sampling schemes is typically guaranteed with high probability by selecting entries to observe uniformly at random from a dense grid \cite{MC,MC2,MC3,MC4}. This is difficult to achieve in seismology since it corresponds to sources and receivers removed uniformly at random from equipment arrays independently for each survey. 

This incompatibility between theory and practice further complicates step 4) for non-Nyquist-based acquisition design. Such a process would involve simulating a 5D data tensor (corresponding to a 3D seismic survey) and performing large-scale reconstruction for all possible options within physical constraints, e.g., solving the \emph{nuclear norm minimization} problem \cite{fazel}. As a consequence, LRMR and CS-based acquisition design is a complex procedure that requires computational exploration of numerous criteria to decide upon an appropriate source-receiver layout. 

To mitigate such tedious processes, we propose a computationally cheap metric to evaluate a given acquisition design. The novel idea comes from the matrix completion literature \cite{UMC,dMC,dMC2}, where the authors show that matrix completion is successful if the \textit{spectral gap} (SG) of the associated set of observed entries is large. To elaborate, let $\textbf{X}\in\mathbb{C}^{n\times m}$ represent densely sampled seismic data (in some transform domain) reshaped as a matrix that exhibits \textit{low-rank structure}, i.e., $\textbf{X}$ is well approximated (according to some tolerance) by a rank-$r$ matrix where $r \ll \min(n,m)$ \cite{SMC,SMC3}. Relative to this desired densely sampled data, we subsample and let $\textbf{M}\in\{0,1\}^{n\times m}$ be a binary matrix whose non-zero entries specify the sampled matrix entries. The SG is defined as the difference between $\textbf{M}$'s first and second singular values $\sigma_1(\textbf{M}),\sigma_2(\textbf{M})$. The main results in \cite{UMC,dMC2} state that one can recover an \textit{incoherent} (i.e., ``not spiky'') low-rank matrix via \emph{nuclear norm minimization} if the number of samples and the SG are large enough. Intuitively, the SG can be seen as a qualitative measure informing us how well we can expect a given sampling scheme to perform matrix completion. 

In this work, we propose the SG as a deterministic measure to differentiate appropriate surveys at an initial design stage prior to computationally expensive simulations and in-field acquisition. This paper argues that the SG provides a simple but powerful tool that can help severely lessen the amount of numerical experiments required to evaluate pre-survey field designs (e.g., simulations via data models and large-scale optimization). Our work aims to demonstrate that the SG can be incorporated into survey planning schemes as a numerically cheap and flexible step to reduce the overall design complexity.

\begin{figure}[!htb]
\centering
\captionsetup[subfigure]{labelformat=empty}
\subfloat{\includegraphics[width=0.75\hsize]{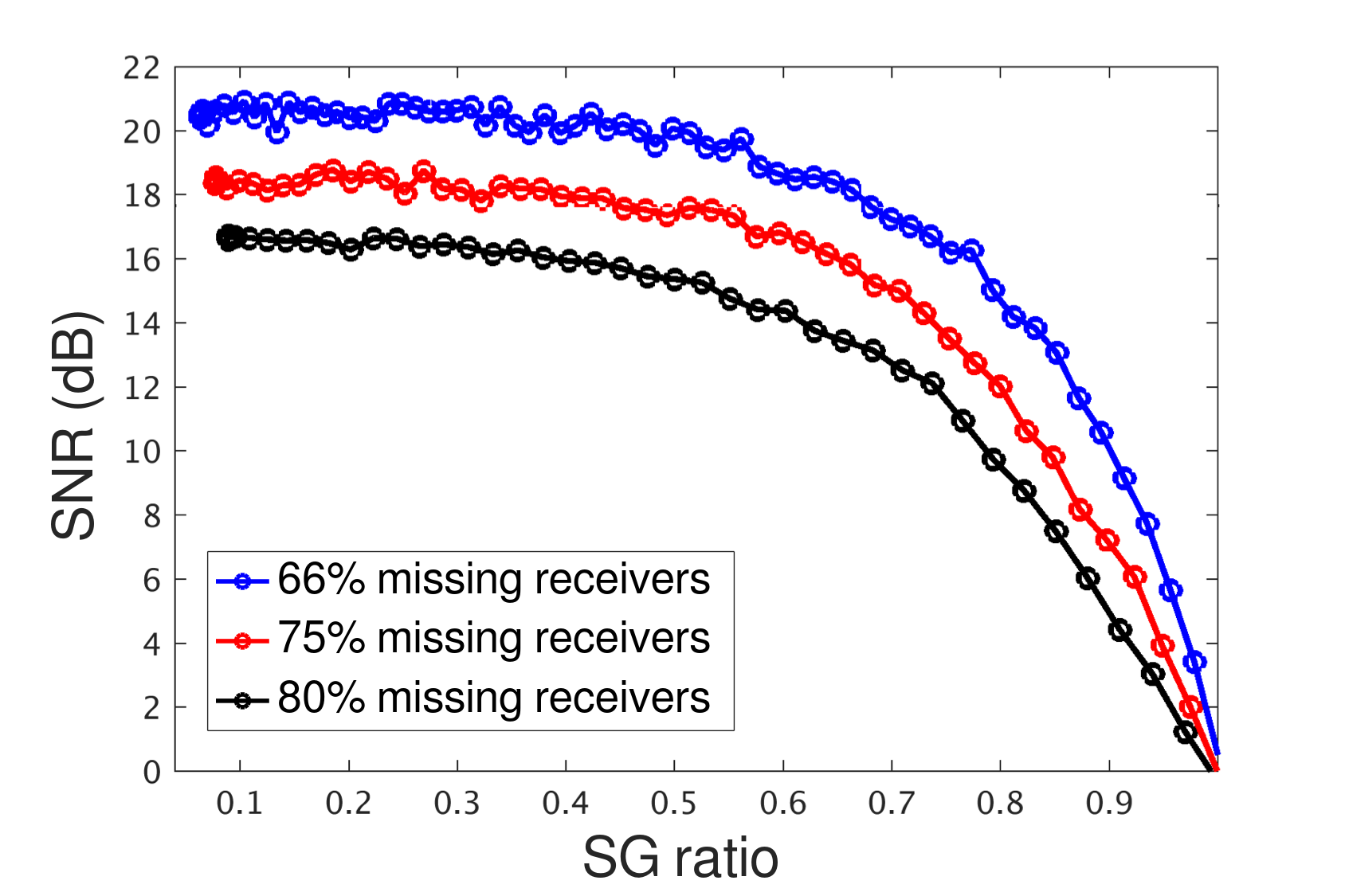}}
\caption{Illustration of the relationship between the quality of reconstruction and the SG ratio of the sampled matrix entries. The figure plots the average reconstruction SNR (defined in (\ref{snr})) vs average SG ratio $\frac{\sigma_2(\textbf{M})}{\sigma_1(\textbf{M})}$ of the corresponding binary sampling matrix $\textbf{M}$, for 3 percentages of missing receivers. The results are taken from the average of 100 independent experiments.}\label{SNRvsSG2}
\end{figure}

The relationship between the SG and quality of reconstruction is demonstrated in Figure \ref{SNRvsSG2}. Given $\textbf{M}\in\{0,1\}^{n\times m}$ specifying the sampled data entries, we constitute its SG by the ratio $\frac{\sigma_2(\textbf{M})}{\sigma_1(\textbf{M})}\in[0,1]$. Henceforth, we will refer to the term $\frac{\sigma_2(\textbf{M})}{\sigma_1(\textbf{M})}$ as the \emph{SG ratio} and note that a large SG corresponds to a small SG ratio ($\frac{\sigma_2(\textbf{M})}{\sigma_1(\textbf{M})}\ll 1$) and vice versa. Figure \ref{SNRvsSG2} illustrates the correlation between $\textbf{M}$'s SG ratio and the signal-to-noise ratio (SNR, defined in (\ref{snr})) of reconstruction via LRMR with observed matrix entries according to $\textbf{M}$. In each experiment, with fixed sampling percentage, only the distribution of the observed entries and corresponding SG change. Therefore, the plot showcases better quality of reconstruction from samples with larger SG. The full details of the conducted experiments are postponed until Section \ref{vs}. Further details for generating $\textbf{M}$ from a survey design are discussed in Section \ref{LRMRseismic}.

Design of optimal acquisition geometries suitable for CS is an ongoing area of research with several solutions in the literature relevant to our work. In \cite{mosher}, the authors propose a reconstruction based design procedure where a greedy method outputs an appropriate sensing matrix. The methodology requires data interpolation at each step of the optimization problem, thus making it a computationally demanding process. The work in \cite{nihed} uses prior information from the low-frequency spectrum of seismic data to construct a signal and coherent noise model, followed by the evaluation of a quality metric for acquisition layout as a function of frequency. The optimal acquisition layout corresponds to the worst case value of the metric over the band. The proposed method does not require any synthetic or real data, but the metric involves computing several condition numbers (i.e, ratio of the smallest and largest singular values of the sensing matrix). This ratio is numerically expensive for sensing matrices related to large-scale 5D seismic data. In contrast, our proposed method requires computing the two largest singular values of a single binary matrix representing source-receiver layout. This approach is numerically efficient and does not require underlying data nor a prior model.  

We validate the SG metric with several numerical experiments that consider distinct aspects of the seismic data acquisition and reconstruction process. In the next section, we begin with some background from the literature of matrix completion that formally introduces the SG. Therein, Section \ref{LRMRseismic} gives a detailed description of how the spectral gap can be computed from a survey design. Section \ref{numexp} is dedicated to numerical experiments, where in Section \ref{vs} we empirically demonstrate the relationship between the SG and quality of reconstruction via LRMR. Section \ref{JSG} demonstrates how the SG is sensitive to large gaps of missing data and in Section \ref{CSM} we utilize the SG to determine an appropriate density of the output interpolation grid. Section \ref{sparsity} argues that the SG is also informative in sparsity-based recovery. We end with concluding remarks and future work in Section \ref{conclusion}.

\section{Universal and Deterministic Matrix Completion}
\label{UMC}
In contrast to most results in the matrix completion literature, the work in \cite{UMC,dMC,dMC2} aims for \textit{universal} and deterministic recovery guarantees. The term universal refers to the set of observed matrix entries, $\Omega\subset \{1,2,\cdots,n\}\times \{1,2,\cdots,m\}$, where this subset is expected to provide an accurate output via nuclear norm minimization for an entire class of signals uniformly. Such results are of great value to practitioners since $\Omega$ will function for all matrices of interest, removing the need to design new sampling schemes for every independent seismic survey. Furthermore, the attached conditions that ensure the suitability of $\Omega$ inherently provide a means to quantify the source-receiver layout. 

This last observation is the focus of this article, where the spectral gap serves as a computationally cheap step to distinguish between given acquisition designs. The proposed metric differentiates sampling patterns in terms of their suitability for reconstruction via matrix completion. When incorporated into existing survey design schemes, the SG offers an important tool that can offset numerically expensive simulation-based acquisition design.

To elaborate, we begin with some matrix completion background. In particular, we focus on deterministic and universal matrix completion \cite{UMC,dMC,dMC2} where the authors quantify the success of matrix completion via the spectral gap. In what follows, let $n\geq m$ without loss of generality. 

Given a matrix of interest $\textbf{X}\in\mathbb{C}^{n\times m}$ and a subset of observed entries $\Omega\subset \{1,2,\cdots,n\}\times \{1,2,\cdots,m\}$, our collected data is given as $\textbf{B} = P_{\Omega}(\textbf{X}+\textbf{E}) \in\mathbb{C}^{n\times m}$, where for $(k,\ell)\in \{1,2,\cdots,n\}\times \{1,2,\cdots,m\}$
\begin{equation}
P_{\Omega}(\textbf{X}+\textbf{E})_{k\ell} =
\left\{
	\begin{array}{ll}
		\textbf{X}_{k\ell} + \textbf{E}_{k\ell} & \mbox{if } (k,\ell)\in\Omega \\
		0 & \mbox{otherwise }
	\end{array}
\right.
\end{equation}
and $\textbf{E}$ encompasses the noise in our observations with bounded Frobenius norm
\[
\|P_{\Omega}(\textbf{E})\|_F := \left(\sum_{(k,\ell)\in \Omega}|\textbf{E}_{k\ell}|^2\right)^{1/2} \leq \eta. 
\]
Matrix completion via nuclear norm minimization aims to approximate
\begin{equation}
\label{nucnorm}
\textbf{X} \approx \textbf{X}^{\sharp} := \underset{\textbf{Z}\in\mathbb{C}^{n\times m}}{\text{argmin}}\quad\|\textbf{Z}\|_* \ \ \mbox{s.t.} \ \ \|P_{\Omega}(\textbf{Z})-\textbf{B}\|_F \leq \eta,
\end{equation}

where
\[
\|\textbf{Z}\|_* = \sum_{k=1}^{m}\sigma_{k}(\textbf{Z})
\]
is the \emph{nuclear norm} and $\sigma_{k}(\textbf{Z})$ denotes the $k$-th largest singular value of $\textbf{Z}$. This method has been extensively studied \cite{MC,MC2,MC3,MC4}. In the case $\eta = 0$, standard results in the literature show that $\textbf{X}$ can be exactly recovered via (\ref{nucnorm}) with high probability under conditions of the general form: 
\begin{itemize}
\item rank$(\textbf{X}) = r \leq \min\{n,m\}$. 
\item $\textbf{X}$ is incoherent, e.g., satisfies the strong incoherence property (see \cite{MC,MC2,MC3,MC4,UMC}). Intuitively, this assumption ensures that the energy (i.e., Frobenius norm) of the data matrix is not concentrated on a small subset of entries.
\item $\Omega$ is generated uniformly at random with $|\Omega| \sim nr \log^{2}(n)$, where $|\Omega|$ indicates the number of observed entries. Furthermore, each matrix to be recovered requires an independent $\Omega$ to be generated.
\end{itemize}
The first condition is crucial since equation (\ref{nucnorm}) with $|\Omega|< nm$ is an underdetermined problem and the low-rank constraint aims to compensate for the limited number of observations. This validates the use of the nuclear norm as the objective function, in order to output $\textbf{X}^{\sharp}$ that can be written as a sum of few rank 1 matrices (i.e., low-rank structure). The second assumption is a necessary condition for sampling schemes that are oblivious of the matrix structure, but can be avoided if one samples according to prior knowledge of the matrix's leverage scores \cite{LMC}. 

However, the third condition is restrictive and does not hold in many applications. For example, in seismic acquisition the allowed structure of $\Omega$ is limited by measurement hardware and physical constraints involved in data collection. Achieving uniform random sampling independently for each survey would require a significant amount of novel sensing equipment. Instead, matrix completion results that consider achievable sampling schemes would be most informative and applicable.

To this end, we now consider results involving universal and deterministic matrix completion \cite{UMC,dMC,dMC2}. Let $\textbf{1}$ be the all-ones $n\times m$ matrix and $\textbf{M} := P_{\Omega}(\textbf{1})\in \{0,1\}^{n\times m}$. We will refer to $\textbf{M}$ as the \textit{sampling mask} and note that it is a binary matrix, which specifies the locations of sampled matrix entries. The main result in \cite{UMC} shows that if the spectral gap (SG) of $\textbf{M}$ is large enough (i.e., $\sigma_2(\textbf{M})\ll\sigma_1(\textbf{M})$) then one can recover any low-rank incoherent matrix via (\ref{nucnorm}). Specifically, the result reads as follows:

\begin{theorem}[Theorem 4.2 in \cite{UMC}]
\label{spectralgap}
Let $\textbf{X}\in \mathbb{C}^{n\times m}$ be a rank-$r$ matrix satisfying the strong incoherence property with parameter $\mu$ (see \cite{MC4} and Claim 5.1 in \cite{UMC}). Let $\Omega$ be generated such that the top singular vectors of $\textbf{M}$ are the all 1's vectors and 
\[
|\Omega| \geq 36\frac{\sigma_2(\textbf{M})^2}{\sigma_1(\textbf{M})}\mu^2 nr^2. 
\]
Then $\textbf{X}$ is the unique optimum of (\ref{nucnorm}) with $\eta = 0$.
\end{theorem}

In the result, the multiplicative factor 36 is an absolute constant (not necessarily optimal) obtained by the authors, independent of the matrix dimensions. This factor bears no particular significance and can likely be improved to further reduce the number of samples needed for universal matrix completion. The incoherence parameter $\mu$ quantifies how evenly spread the information is throughout the data matrix and ensures that we sample accordingly. The condition on the singular vectors of $\textbf{M}$ requires $\Omega$ to contain the same number of samples in each row and column. This constraint is quite restrictive, but also required for the main results in \cite{dMC,dMC2}. Generating sampling schemes with this structure is not straightforward \cite{graph2,graph3} and complex to translate to field samples. However, in Section \ref{numexp}, we allow ourselves the liberty to violate this assumption. Our numerical experiments will demonstrate empirically that the relationship between the SG and the success of matrix completion remains useful even when this condition does not hold.

In contrast to standard uniform random sampling requirements, $|\Omega|\sim\mathcal{O}(nr\log^{2}(n))$, Theorem \ref{spectralgap} requires $|\Omega|\sim\mathcal{O}(nr^2)$. While this sample complexity is pessimistic when $r>\log^{2}(n)$, we stress that the result holds deterministically and universally for all incoherent matrices. In the best case scenario we can expect $\frac{\sigma_2(\textbf{M})^2}{\sigma_1(\textbf{M})} \sim \mathcal{O}(1)$ (e.g., Ramanujan graph \cite{Ramanujan,UMC}). More importantly, Theorem \ref{spectralgap} provides a useful relationship between the success of matrix completion and the ratio $\frac{\sigma_2(\textbf{M})}{\sigma_1(\textbf{M})}\in [0,1]$ directly related to the SG. This ratio appears with similar implications in the related work \cite{dMC,dMC2}. Intuitively, a practitioner choosing amongst several sampling designs can compute the respective SG ratios and make an educated choice of which design will output the best quality of matrix reconstruction.

\subsection{LRMR-Based Seismic Data Acquisition}
\label{LRMRseismic}

Our work proposes the SG as a useful metric for survey planning that differentiates sampling schemes at varying stages of existing designs. Keeping in mind the results from the previous section, our proposed matrix completion-based seismic data acquisition and reconstruction hinges upon the following three requisites: 
\begin{enumerate}[i)]
\item Densely sampled seismic data (in some transform domain) reshaped as a matrix $\textbf{X}\in\mathbb{C}^{n\times m}$ should exhibit incoherence and low-rank structure, i.e., $\textbf{X}$ is well approximated (according to some tolerance) by a rank-$r$ matrix where $r \ll \min(n,m)$.
\item A computationally efficient program for large-scale LRMR, e.g., nuclear norm minimization \cite{fazel}. 
\item The subset of observed matrix entries $\Omega\subset \{1,2,\cdots,n\}\times \{1,2,\cdots,m\}$ should be suitable for incoherent matrices.
\end{enumerate}

To address i), we note that several reorganizations and transformations of seismic data tensors into incoherent matrices have been shown to expunge low-rank structure \cite{SMC,SMC3}. Furthermore, for ii) efficient methodologies have been proposed to solve the matrix completion problem with large matrices \cite{slim,RCAM,AMC}, including on parallel architectures \cite{PMC,NOMAD}. 

This paper focuses on fulfilling iii) in a practical manner using the SG. While specific circumstances will vary, our general approach consists of generating a sampling mask $\textbf{M}$ for each acquisition under consideration and comparing them via the SG. To do so, a practitioner must first determine the desired output sampling density as in steps 1) and 2) from Section \ref{introduction}. This will implicitly dictate the matrix dimensions $n$ and $m$, which also depend on the transform domain chosen to satisfy requisite i) above. A given sampling scheme will sub-sample relative to this desired source-receiver spacing, i.e., collect wavefield measurements on a subset of $\{1,2,\cdots,n\}\times \{1,2,\cdots,m\}$ and thereby determine $\Omega$. The sampling mask is then defined as the $n\times m$ matrix with value of 1 on the entries in $\Omega$ and 0's otherwise. Finally, the singular values $\sigma_1(\textbf{M}),\sigma_2(\textbf{M})$ and the respective gap (or SG ratio) can be computed.

Comparing the SG ratios of each acquisition design considered provides a computationally efficient mechanism to select or eliminate sub-sampling schemes based on their suitability for matrix completion. This procedure can alleviate the complexity of simulation-based evaluations for survey design, where the SG can reduce the number of computationally intense experiments typically required for steps 3) and 4) from Section \ref{introduction}. 

It is important to notice that in the context of seismology, the SG is oblivious to the underlying geological complexity. Therefore, the SG tool proposed here does not consider many important attributes of seismic survey design related to specific models or geological scenes (e.g., evaluating azimuth coverage and fold \cite{spread,azimuthref}). Such topics are an important component of standard acquisition design \cite{SeisAcqBook}, and should be addressed in conjunction. In this sense, our proposed SG metric should not be considered sufficient for a complete survey design but rather as a tool that can reduce the numerical complexity of existing workflows. In Section \ref{numexp}, we demonstrate how the SG can be used to limit the amount of data-dependent evaluations and how it largely succeeds at evaluating surveys under models of varying geological complexity.

\subsection{Connections with Graph Theory}
Interestingly, the SG is a common tool for analyzing the connectivity of communication networks \cite{Ramanujan,netgap}. In this context, a larger SG provides a more robust network. The foundation comes from graph theory and the study of expander graphs, where a communication network can be seen as a group of vertices (network nodes) with corresponding edges (signifying communication between nodes). Considering the respective adjacency matrix (analogous to $\textbf{M}$ in our context), this matrix's SG gives a direct measurement of the network's efficiency in the flow of information. 

A similar interpretation can be presented in our seismology scenario. The array of sources and receivers can be seen as the vertices with edges that represent the recording and transmission of pressure waves. In other words, using the matricization in Section \ref{vs} (see also \cite{matrici,SMC}), a source$_x$-receiver$_x$ pair (vertex in the $x$-axis) will have an edge with a source$_y$-receiver$_y$ pair (vertex in the $y$-axis) if the receiver$_x$-receiver$_y$ pair recorded the pressure wave due to the corresponding source$_x$-source$_y$ pair (and vice versa with respect to the reciprocal role of transmission). Thus, in our case the sampling mask $\textbf{M}$ can be seen as the corresponding biadjacency matrix of the bipartite graph that captures a sampling scenario with such vertices and edges. Since the interpolation problem implicitly depends on well distributed samples, we can intuitively see that a well connected graph (i.e., large SG) will be more effective at infilling data between source-receiver pairs. This interpretation was first elaborated abstractly in \cite{MC2}, where the authors show that observed entries generating a fully connected graph are necessary to recover a matrix of any rank.

\section{Numerical Experiments}
\label{numexp}
To demonstrate the utility of the SG for survey design, we numerically address four specific questions related to seismic data acquisition and reconstruction: 
\begin{itemize}
\item Section \ref{vs}: What is the relationship between the SG and the quality of LRMR-based reconstruction?
\item Section \ref{JSG}: Is the SG sensitive to large gaps of missing data?
\item Section \ref{CSM}: Can the SG identify appropriate output interpolation grid density?
\item Section \ref{sparsity}: Is the SG informative for sparsity-based reconstruction?

\end{itemize}

We design numerical experiments to provide insight for these questions. Some of our experiments are abstract in the sense that we do not consider explicit seismic data, and instead focus on the source-receiver distribution. This stresses that the SG can be applied prior to data collection and without simulated data. However, most industrial acquisition designs require simulation on data models \cite{verschuur,modelsurvey}. To demonstrate the utility of the SG in such workflows, we also provide experiments with validation on geologically diverse synthetic marine datasets (BG 3D, SEG/EAGE Overthrust and the SEAM models). 

These experiments aim to demonstrate that even though the SG is oblivious to the underlying data, the reconstruction quality predicted by this tool largely pans out for models of varying geological complexity. While these examples do not prove the proposed method will always work, they indicate that the spectral gap is a good metric for survey suitability that can be incorporated into standard simulation-based design methodologies. Furthermore, we note that our experiments mainly focus on seismology in marine settings. Land surveys require distinct considerations and separate experiments are needed to fully explore the role of the SG in such acquisitions. However, due to its detachment from the underlying data, we believe that the experiments here will be informative for the applicability of the SG in land acquisition geometries.

In what follows, we use the signal-to-noise ratio (SNR) as a metric of matrix reconstruction quality defined as
\begin{equation}
\label{snr}
-20\log_{10}\left(\frac{\|\textbf{X}^{\sharp}-\textbf{X}\|_F}{\|\textbf{X}\|_F}\right)
\end{equation}
where $\textbf{X}\in\mathbb{C}^{n\times m}$ is the desired data matrix, $\textbf{X}^{\sharp}\in\mathbb{C}^{n\times m}$ is the reconstructed approximation.
%

\subsection{Relationship between the SG and quality of reconstruction}
\label{vs}
\begin{figure}[!htb]
\centering
\captionsetup[subfigure]{labelformat=empty}
\subfloat[a]{\includegraphics[width=0.7\hsize]{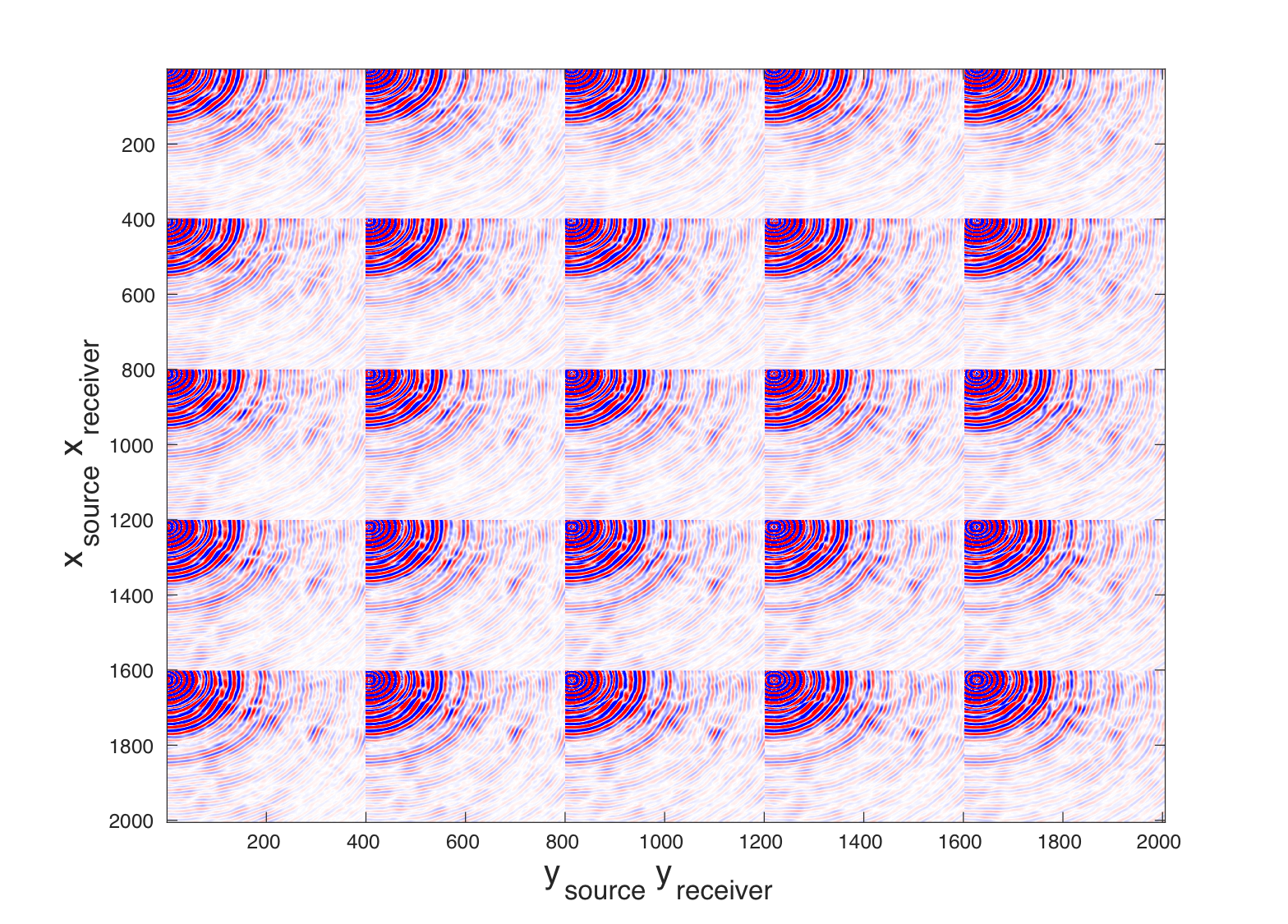}}\\
\subfloat[b]{\includegraphics[width=0.7\hsize]{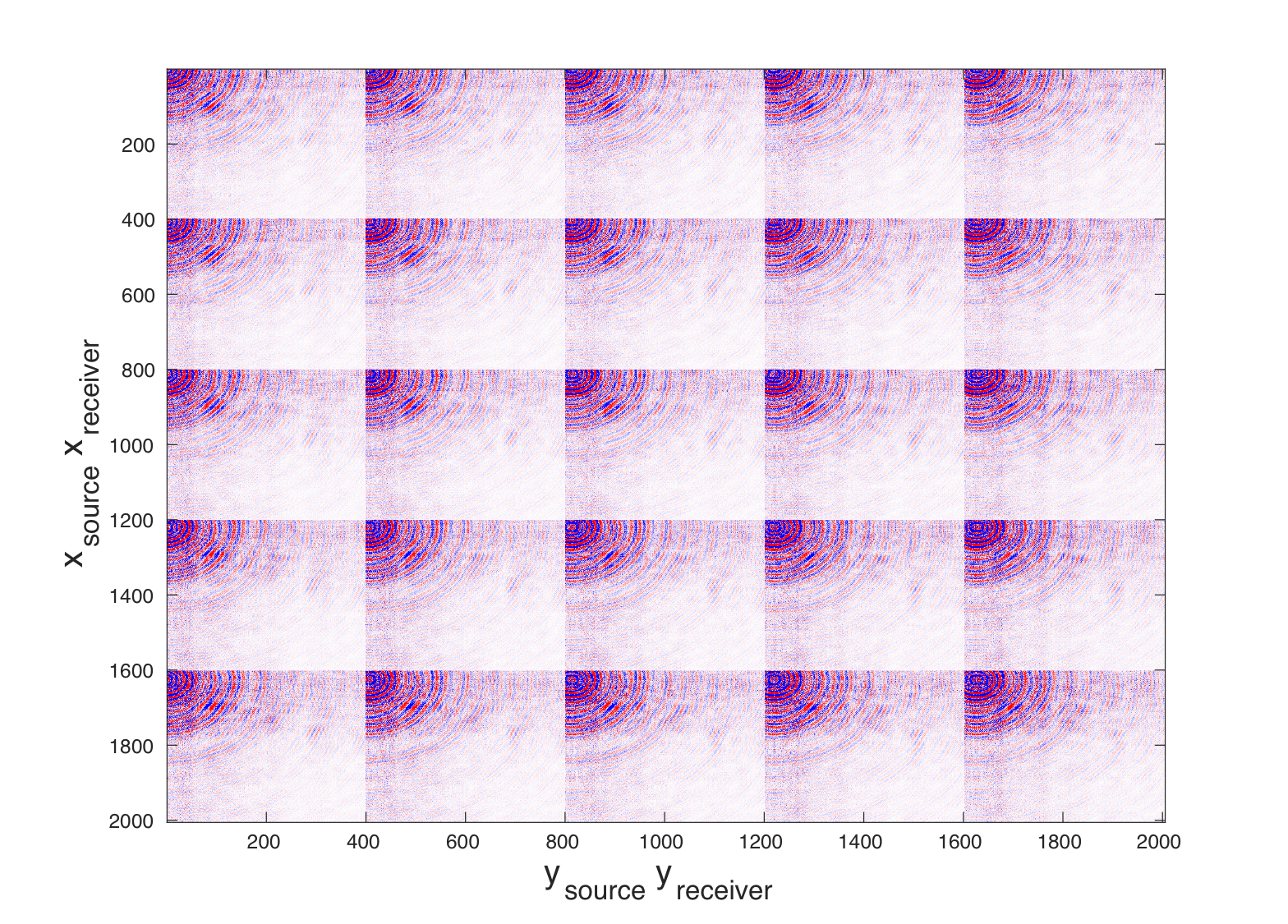}}\\
\subfloat[c]{\includegraphics[width=0.7\hsize]{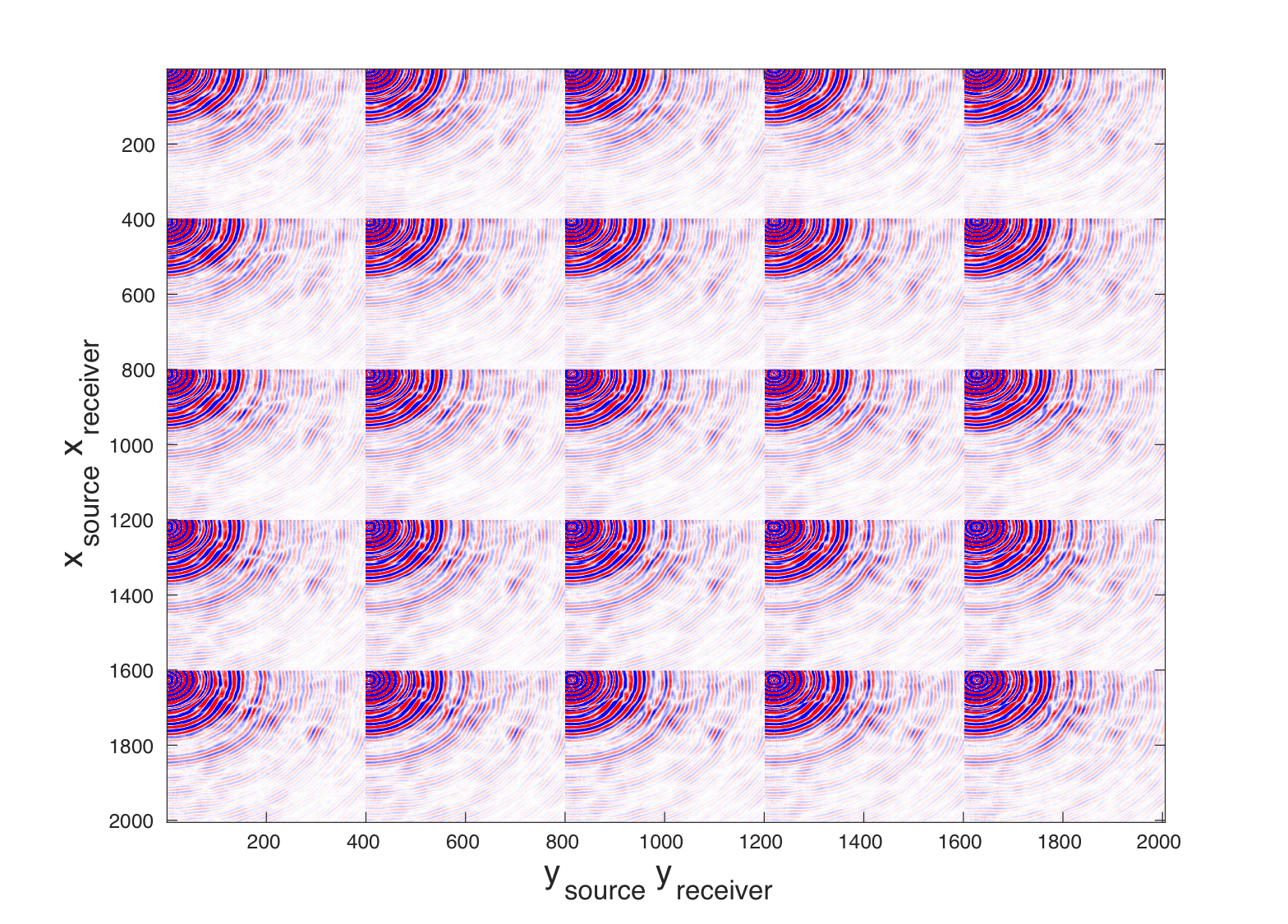}}
\caption{Examples of recovered data matrices via sampling masks with small and large SG. (a) True data (b) reconstructed data with sampling mask exhibiting SG ratio $= .9828$ and recovery SNR $= 3.5$ dB ($75 \%$ missing receivers), and (c) reconstructed data with sampling mask exhibiting SG ratio $= .1796$ with SNR $= 20.7$ dB ($75 \%$ missing receivers).}\label{SNRvsSG}
\end{figure}

We begin with examples that numerically demonstrate the relationship between the SG and the quality of LRMR output. We perform experiments on 5D data tensors (corresponding to 3D seismic surveys respectively). We consider the BG 3D model with 68 sources and 401 receivers along each axis to generate 5D data tensor with dimensions: time, source$_x$, source$_y$, receiver$_x$, and receiver$_y$. We work in the Fourier domain by applying a fast Fourier transform (FFT) along the time axis and reconstruct in a frequency by frequency basis (generating a sequence of 4D tensors). We use the matricization from \cite{SMC} (first introduced in \cite{matrici}) to reorganize each 4D tensor into a matrix, where the source$_x$-receiver$_x$ pairs are coupled along the rows and the source$_y$-receiver$_y$ pairs are coupled along the columns (with the sources serving as the outer dimensions). Henceforth, we will refer to this matricization as \emph{src-rec form}. We restrict ourselves to a single 7.34 Hz frequency slice and the first 5 sources along each axis, generating a $2005\times 2005$ matrix (see Figure \ref{SNRvsSG}a). 

We begin by removing a specified number of receivers (according to a sampling percentage) in a periodic manner (e.g., keep only first receiver of every three for approximately $66\%$ missing receivers). We then solve the nuclear norm minimization problem (\ref{nucnorm}) with $\eta = 0$, applying the corresponding sampling mask and compute its SG ratio $\frac{\sigma_2(\textbf{M})}{\sigma_1(\textbf{M})}$. Subsequently, we choose a specified percentage $p\in[0,1]$ of the observed receivers and relocate them among the unobserved receiver locations in a uniform random manner. This generates sampling masks with the specified sampling percentage and varying levels of randomly placed receivers. We now solve the nuclear norm minimization problem with the modified sampling mask and record the respective SG ratio. We repeat this procedure for several percentage $p$ values in an increasing manner (specifically, $p \in \{0,.02,.04.,\cdots,1\}$).

In general, we observe a correlation between the number of randomly placed receivers and the SG of the resulting sampling mask, where masks with larger degrees of randomness provide larger SG values (i.e., smaller SG ratios $\frac{\sigma_2}{\sigma_1}$). This whole procedure is repeated 100 times independently and we average the SG ratios and SNR values for each $p \in \{0,.02,.04.,\cdots,1\}$. To obtain Figure \ref{SNRvsSG2} we sort the resulting SG ratios in increasing order and plot them against the corresponding SNR of reconstruction. Furthermore, Figures \ref{SNRvsSG}b and \ref{SNRvsSG}c show two recovered data sets from sampling masks with small and large SG, respectively. These figures imply a clear correlation between the SG and the SNR of reconstruction.

Notice that the SNR values in Figure \ref{SNRvsSG2} taper off rather quickly towards the optimal SNR value in each case. This observation implies that a practitioner need not place all receivers randomly and might achieve a similar quality of reconstruction with a hybrid layout (e.g., half periodic and half random receiver placement). Such acquisition designs provide achievable surveys with near-optimal SG ratios (i.e., close to those of a Ramanujan graph \cite{Ramanujan}), where the resulting survey minimizes the amount of receivers not placed in a convenient and patterned fashion.

\subsection{Sensitivity of the SG to large gaps of missing data}
\label{JSG}
The previous section implicitly demonstrated the utility of randomness for LRMR appropriate sampling via the SG. However, such random samples may exhibit large gaps of missing data that incur degrading artifacts on many interpolation techniques \cite{gaps}. In this section, we explore the sensitivity of the SG to such gaps. For this purpose we use \textit{jittered subsampling} \cite{jitter}, i.e., we partition the desired dense receiver grid into intervals with equal number of receivers and randomly choose a single receiver location to sample from each interval (where the number of receivers per interval depends on the sampling ratio). These sampling techniques are common in computer graphics to attenuate aliasing artifacts while controlling the gap size of unrendered data \cite{pixar}.

We design an abstract 3D survey (corresponding to a 5D data tensor) with 32 sources and receivers in each axis and restrict ourselves to a single frequency slice, working in Fourier domain and src-rec form as in the previous section (generating a $1024\times 1024$ matrix). To emphasize that our methodology does not require numerically expensive data simulation, we do not use an explicit dataset here and instead provide our argument via the binary sampling masks only. 

We remove $80\%$ of the receivers via several jitter subsampling methods that generate varying average gap sizes of missing receiver data (here each interval contains 5 receivers). We do so by varying the probability distribution by which we choose the receiver location for every other interval, while the remaining intervals retain the uniform distribution (i.e., each receiver location has equal probability of being chosen). The selection distribution is modified by selecting a \textit{jitter parameter} $:=\rho\in[0,1]$ and only choosing a sample among the first $\lceil\rho\cdot 5\rceil$ receiver locations in the interval. This provides receiver locations with increasing average gap sizes (as $\rho$ decreases). Specifically we consider $\rho \in\{.2,.4,.6,.8,1\}$. By only modifying every other interval's selection distribution we still assure a rich sampling scheme (i.e., each row and column will be sampled with high probability in the chosen matricization).

\begin{figure}[!htb]
\centering
\captionsetup[subfigure]{labelformat=empty}
\subfloat{\includegraphics[width=0.75\hsize]{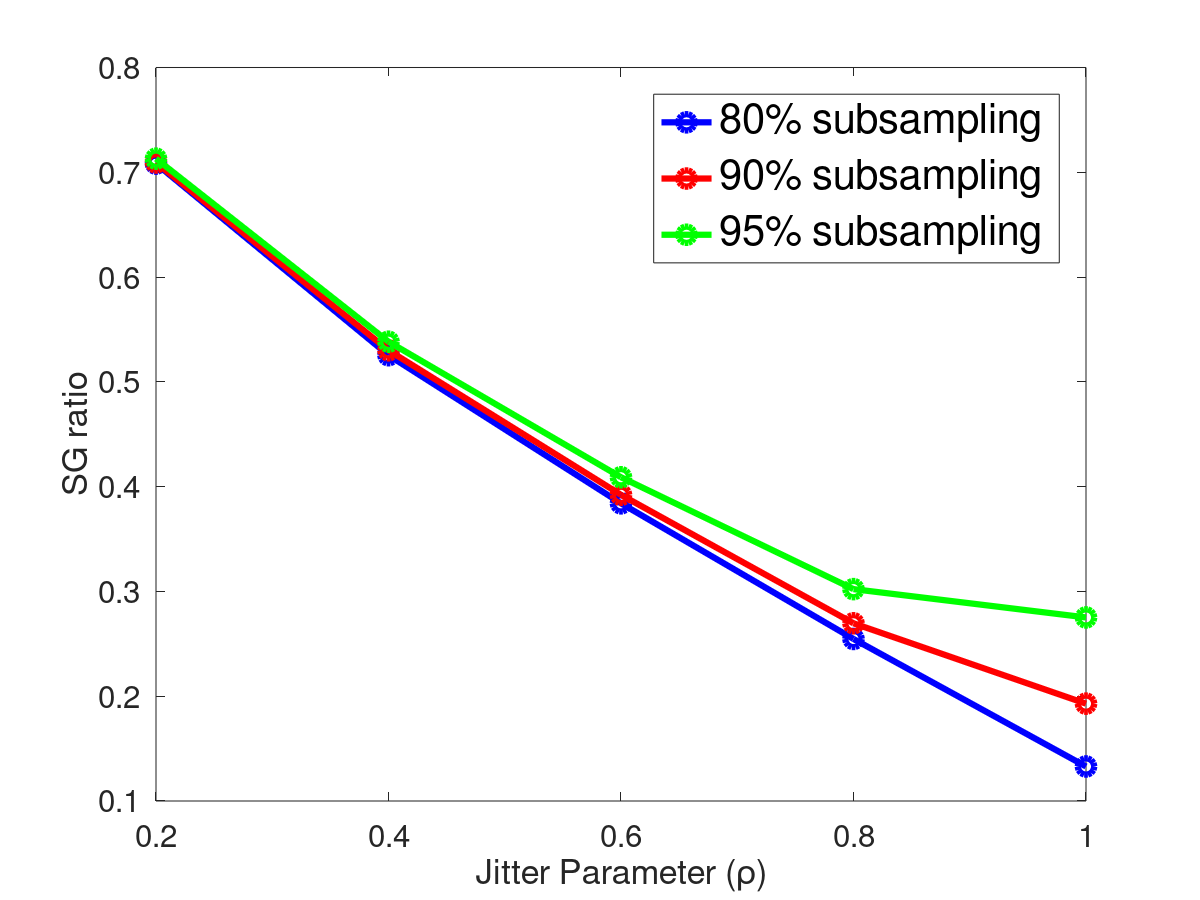}}
\caption{Plot of average SG ratio of sampling mask vs jitter parameter $\rho$, for various sampling percentages. The results are taken from the average of 200 independent experiments. Notice that as the average sampling gap increases ($\rho\rightarrow 0$) we obtain less favorable SG values (and diminished quality of reconstruction as expected). The SG is therefore sensitive to large gaps of missing data and we expect the highest quality of reconstruction via optimally-jittered subsampling ($\rho = 1$).}\label{JitvsSG}
\end{figure}

We compute the SG ratio of each sampling mask generated for each $\rho$ value specified above. We repeat the experiments 200 times and plot the average SG ratio against the corresponding jitter parameter $\rho$ in Figure \ref{JitvsSG}. The experiments are repeated for $90\%$ and $95\%$ missing receivers. Figure \ref{JitvsSG} exhibits a clear trend between the jitter parameter $\rho$ and the SG, where $\rho\approx 1$ provides the largest SG (smallest SG ratios $\frac{\sigma_2}{\sigma_1}$). The case $\rho = 1$ corresponds to \textit{optimally-jittered} subsampling \cite{jitter}, with smallest average gap size between receiver locations. Arguably, this type of jitter is best for general sampling and the SG can successfully make this distinction. The main conclusion of this section is that the SG can be used to detect a well distributed set of samples (which we generate in a random fashion for convenience) while remaining aware of large gaps of missing data. These properties are crucial for universal sampling schemes and the SG is an efficient tool to quantify these in a given set of observed matrix entries.

\subsection{Coil Sampling Experiments}
\label{CSM}

\begin{figure}[!htb]
\centering
\captionsetup[subfigure]{labelformat=empty}
\subfloat{\includegraphics[width=1\hsize]{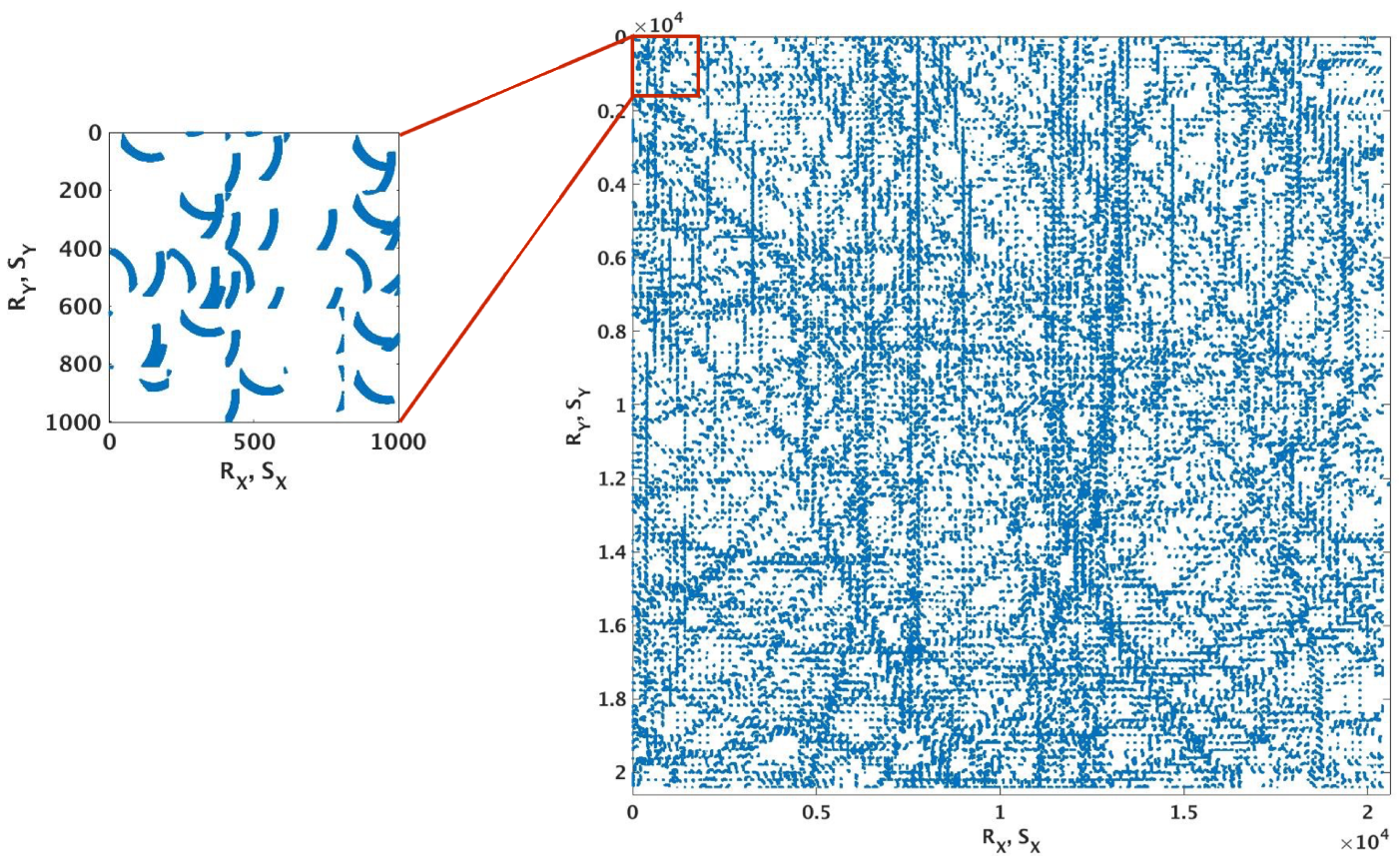}}
\caption{Sampling mask generated from a coil-based marine seismic survey deployed by Schlumberger. The data corresponds to single-coil survey, extracted from a dual-coil survey. The array represents the entire desired densely sampled seismic data reshaped as a matrix in src-rec form (see Section \ref{vs}), where each entry corresponds to a source-receiver pair. Blue entries represent the sampled data, i.e., the receiver$_x$-receiver$_y$ pairs that recorded the pressure wave produced by a source$_x$-source$_y$ pair. }\label{coilmask}
\end{figure}

We consider a 3D marine streamer coil survey, which is designed to have a random distribution of sources and receivers \cite{nick}. For imaging and post-processing, such data sets must be accurately interpolated onto a dense regular grid. We focus on this reconstruction step for a dual-coil streamer data set provided by Schlumberger, deployed on a survey area of $10\times 10$ km (illustrated via src-rec form in Figure \ref{coilmask}). We consider the source-receiver layout indicated by the streamer data, with the goal of choosing an output interpolation grid that is as dense as possible while appropriate for LRMR. The experiments of this section do not utilize the actual pressure wave recordings of the survey (this data remains the property of Schlumberger). Instead, we verify the quality of reconstruction by simulating seismic data using the SEG/EAGE Overthrust model that matches the $10\times 10$ km area from which the coil sampling geometry was extracted. 

We adopt the same work flow from the previous sections, reconstructing sequentially on 4D mono-chromatic tensors matricized via src-rec form and present our results for a single 10 Hz frequency slice. Given a desired spacing of sources and receivers, we create a corresponding sampling mask $\textbf{M}\in \{0,1\}^{n\times m}$, with one's in the entries where streamer data was acquired and zero’s assigned to all other entries. Note that, in the field acquisition, sources and receivers are placed off-the-grid (i.e., not necessarily in our desired output grid). Therefore, to construct the sampling mask we \emph{bin} the data onto the desired reconstruction grid, i.e., we assign each off-the-grid sample to its nearest grid point. We adopt this binning procedure for simplicity in acquisition design, but note that LRMR methodologies exist that incorporate the off-the-grid samples for improved quality of reconstruction \cite{offgrid}.

We examined many output grids by varying the source-receiver density between 50-200 m and showcase three examples in Table \ref{coilexp} along with the respective SG ratios and SNR of reconstruction (computed for the synthetic 10 Hz Overthrust frequency slice). The SG successfully distinguishes which layout will produce the highest quality output in terms of the SNR. Notice that a denser grid is not always optimal since the example in Table \ref{coilexp} with densest 50 m source-receiver spacing exhibits the least desirable SG ratio (and reconstruction SNR). Intuitively, this 50 m grid is relatively too dense for the given coil samples and therefore decreases the sampling percentage with large gaps of missing data that are conveniently detected via the SG.

\begin{table}[!htb]
\centering
\caption{SG \& SNR AS A FUNCTION OF GRID SPACING}
\begin{tabular}{l*{4}{c}r}
Receiver spacing  (m)            & Source spacing (m) & SG ratio ($\frac{\sigma_2}{\sigma_1}$)  & SNR\\
\hline
100 & 200 & 0.77 & 10 dB \\
50            & 100 & 0.4  & 16 dB\\
50           & 50 & 0.88 &  8 dB\\
\end{tabular}
\label{coilexp}
\bigskip

Reconstruction experiments for marine seismic data, where the SG ratios correspond to sampling masks in src-rec form extracted from a dual-coil survey deployed by Schlumberger. The SNR corresponds to LRMR experiments using a simulated 10 hz frequency slice via the SEG/EAGE Overthrust model that matches the survey area.
\end{table}

\subsection{SG for sparsity-based reconstruction}
\label{sparsity}

Although our focus thus far has been on the matrix completion approach, graph spectrum-based techniques are flexible and apply to other signal processing and imaging frameworks (including compressive sensing \cite{graph1}). To validate the SG on sparsity-based seismic data reconstruction, we designed a 2D survey layout for an ocean-bottom node (OBN) acquisition and perform reconstruction using matching pursuit Fourier interpolation \cite{MPFI}. We use synthetic data simulated on the geologically complex SEAM model. The full dataset contains 510 sources and receivers sampled at 10 m. To simulate a realistic subsampling scenario, we sample sources at 50 and 100 m across inline and crossline directions, respectively. This subsampling will cause aliasing to appear from 15 and 7 Hz onwards across the frequency band in the inline and crossline directions. We only perform reconstruction for a common-receiver gather. We interpolate a periodic and a jittered random acquisition design \cite{jitter}, comparing the resulting SG and SNR of reconstruction. 

\begin{figure}[!htb]
\centering
\captionsetup[subfigure]{labelformat=empty}
\subfloat{\includegraphics[width=1\hsize]{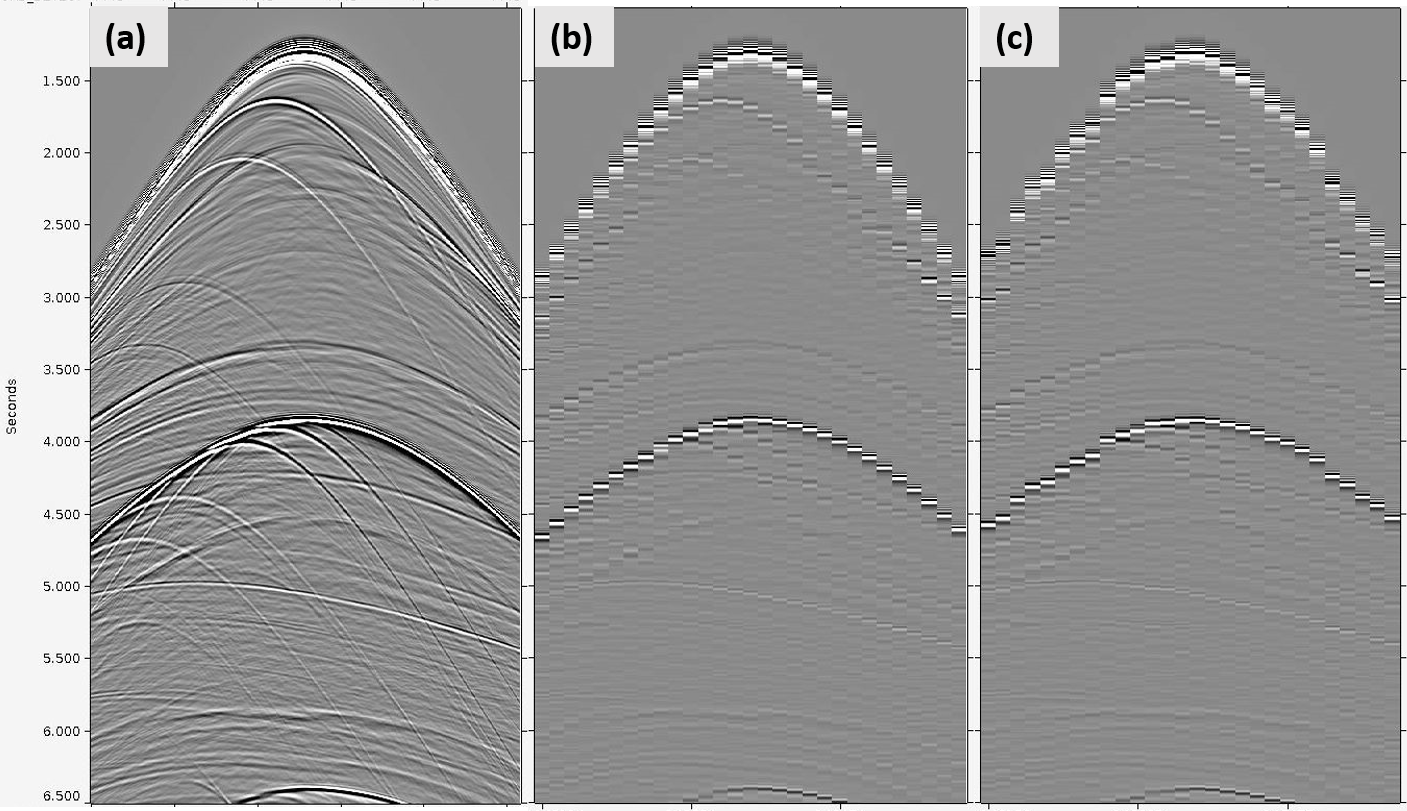}}
\caption{(a) Fully sampled crossline section extracted from the SEAM model. Subsampled crosssection using (b) periodic, and (c) jittered subsampling designs. The SG ratios ($\frac{\sigma_2(\textbf{M})}{\sigma_1(\textbf{M})}$) for (b) and (c) are 1.0 and 0.35, respectivelty.}\label{perjit}
\end{figure}

\begin{figure}[!htb]
\centering
\captionsetup[subfigure]{labelformat=empty}
\subfloat{\includegraphics[width=1\hsize]{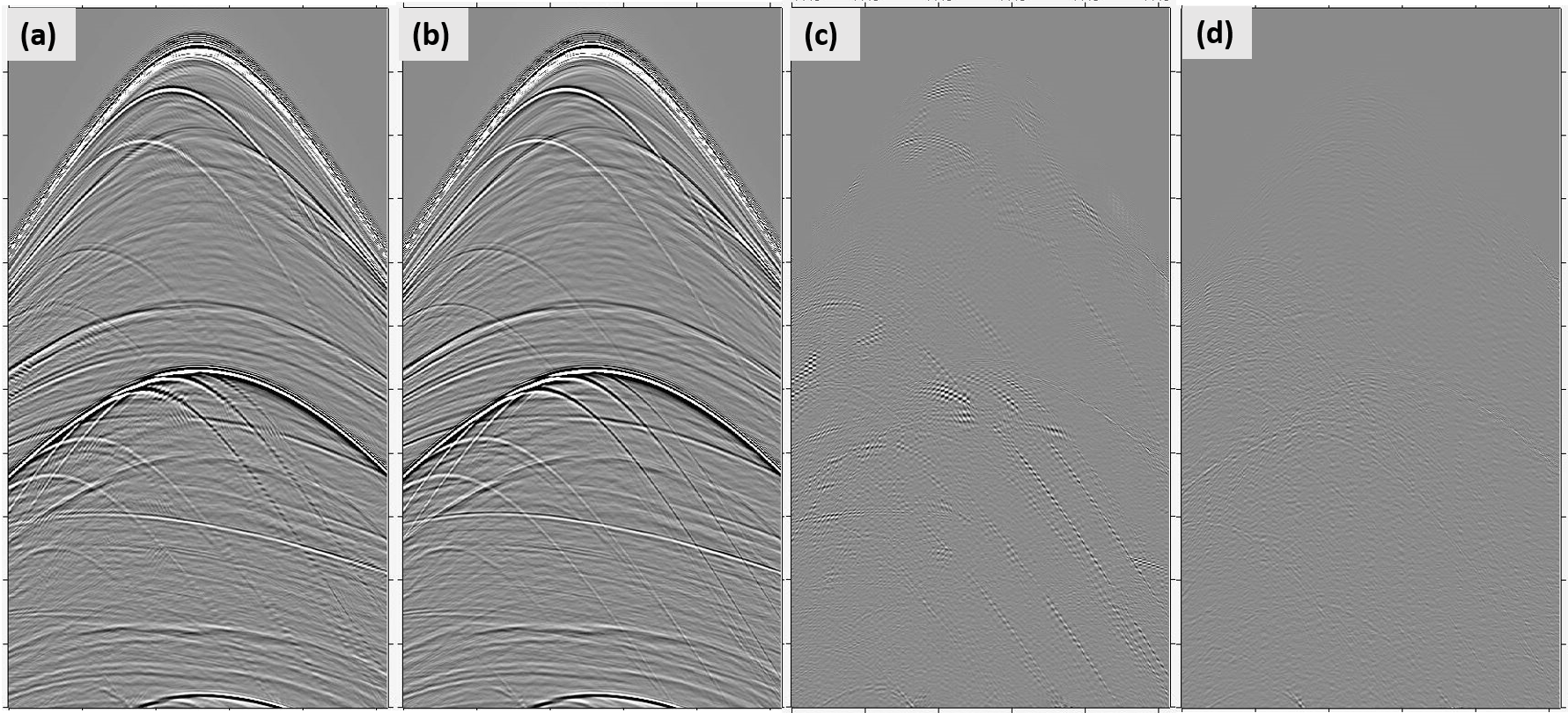}}
\caption{Crossline section from the SEAM model after seismic data reconstruction using (a) periodic, and (b) jittered subsampling, and (c, d) the corresponding residual sections. The reconstruction SNRs for (b) and (c) are 20 dB and 24 dB respectively. In contrast to periodic sampling, we are able to preserve the reflections and diffraction energy well using the jittered subsampling, which is evident in the residual section and the SG.}\label{perjitint}
\end{figure}

The periodic and jittered subsampling masks have respective SG ratios $\frac{\sigma_2(\textbf{M})}{\sigma_1(\textbf{M})}$ of 1 and 0.35. For periodic sampling, we subsample data at every $5^{th}$ and $10^{th}$ grid point along inline and crossline directions. Figure \ref{perjit}a shows the cross-line section from the SEAM dataset, whereas Figures \ref{perjit}b and \ref{perjit}c show the subsampled data using periodic and jittered sampling respectively. Figures \ref{perjitint}a and \ref{perjitint}b show the interpolation results for periodic and jittered subsampling with an SNR values of 20 and 24 dB, respectively. To stabilize the reconstruction, priors were introduced in the data reconstruction framework. The amplitude spectrum derived from the alias-free frequency band of small spatio-temporal windows is incorporated as weights to distinguish between the aliasing artifacts and the true events at the higher frequencies \cite{priora, priorb}. Although the priors significantly improve the results for the periodic sampled data, we still loose coherent energy of complex seismic events. In comparison, the reconstruction via jittered subsampling with priors is able to preserve the continuity of reflection events and complex diffraction patterns. This observation is predicted by the SG and further supported by the residuals in Figures \ref{perjitint}c and \ref{perjitint}d. 


\section{Conclusions and Discussion}
\label{conclusion}
This paper provides a computationally efficient tool that extends the utility of matrix completion and compressive sensing for seismic data acquisition. By considering the work in \cite{UMC,dMC,dMC2}, we propose the spectral gap as a means to select amongst source-receiver layouts in terms of their suitability for reconstruction via low rank matrix recovery and compressive sensing. Our numerical experiments demonstrate the flexibility of this tool and its clear correlation with quality of reconstruction. 

The main advantage of the spectral gap is its relatively low computational complexity, were a practitioner need only compute the first two singular values of a binary matrix. However, the spectral gap is oblivious to the underlying data and does not consider many important aspects of seismic survey design (e.g., azimuth coverage and fold). Because of this, we warn the reader that the spectral gap does not provide a comprehensive survey design by itself. Instead, as demonstrated in this work, the spectral gap can be incorporated into standard survey design workflows. When utilized prior to in field surveys and large-scale optimization, the spectral gap is a flexible tool that severely reduces the amount of simulation-based evaluations required for acquisition design.

Several directions for future work remain. The main results in \cite{UMC,dMC,dMC2} require that $\Omega$ be generated from a regular graph (this is the condition on the singular vectors of the sampling mask $\textbf{M}$ in Theorem \ref{spectralgap}), which may be too restrictive. While our experiments have illustrated that the spectral gap is informative even when this condition may not hold, a general result that relaxes this requisite would be most instructive for cost-effective and low-environmental impact acquisition design. 

Throughout, we implicitly assume that the samples are on-the-grid since our sampling masks are generated by choosing sources-receivers from a dense equispaced grid. However, in practice this is rarely the case since samples often deviate from the desired grid points and in general lie on a continuous domain (as in the dual-coil survey of Section \ref{CSM}). Therefore, coil sampling and other randomized acquisition designs would benefit from an analogous tool that quantifies these off-the-grid source-receiver layouts. 

This work has shown that the spectral gap can be a valuable tool when choosing among a finite set of acquisition designs. Although this alleviates survey design, practitioners must exhaustively consider many abstract sampling schemes to be compared via the spectral gap. An improved tool could utilize the concepts of this work to choose among an infinite set of layouts. This could be done, for example, by setting up a tractable optimization program that finds the sparsest mask with largest spectral gap among all masks that satisfy constraints specified by a desired sampling geometry.

Beyond seismology, one interesting application of the spectral gap could be for designing new 2D medical ultrasound probes. The total number of transducers in a probe is a function of the sampling rate based on Nyquist criteria and the required size of the probe. Technical challenges related to physical wired connections of a large number of transducers makes it necessary to reduce this number. This reduction is non trivial, since the quantity of transducers should not affect the quality of the images generated from probe measurements. Random distribution of transducers is beneficial for improving the ultrasound images \cite{Cobbold}. As the modelling of a large number of transducers is time consuming, the spectral gap could speed up the probe design process.

\section{Acknowledgments}
The authors would like to thank Schlumberger WesternGeco for their support and for providing of the dual-coil streamer data.

\end{document}